\documentclass[twocolumn,amsmath,showpacs,amsfonts,aps,prc,floatfix]{revtex4}
\usepackage{graphicx}
\usepackage{bm}

\begin{document}

\title{Viscous hydrodynamics description of $\phi$ meson production in Au+Au and Cu+Cu collisions}
 
\author{A. K. Chaudhuri}
\email[E-mail:]{akc@veccal.ernet.in}
\affiliation{Variable Energy Cyclotron Centre, 1/AF, Bidhan Nagar, 
Kolkata 700~064, India}

\begin{abstract}

In the Israel-Stewart's theory of 2nd order dissipative hydrodynamics, we have simulated $\phi$ production in Au+Au and Cu+Cu collisions at $\sqrt{s}_{NN}$=200 GeV.  Evolution of QGP fluid with viscosity over the entropy ratio $\eta/s$=0.25, thermalised at $\tau_i$=0.2 fm, with initial energy density $\varepsilon_i$=5.1 $GeV/fm^3$ explains the   experimental data on $\phi$  multiplicity,  integrated $v_2$,  mean $p_T$, $p_T$ spectra and elliptic flow in central and mid-central Au+Au collisions. $\eta/s$=0.25 is also consistent with centrality dependence of $\phi$ $p_T$ spectra in Cu+Cu collisions. The central energy density in Cu+Cu collisions is $\varepsilon_i$=3.48 $GeV/fm^3$.
   
 \end{abstract}

\pacs{47.75.+f, 25.75.-q, 25.75.Ld} 

\date{\today}  

\maketitle


\section{Introduction}
\label{intro}
 
Experiments in Au+Au collisions at RHIC  \cite{BRAHMSwhitepaper,PHOBOSwhitepaper,PHENIXwhitepaper,STARwhitepaper}, produced convincing evidences that in non-central Au+Au collisions, a hot, dense, strongly interacting,  collective QCD matter is created. Whether the matter can be characterized as the lattice QCD \cite{lattice,Cheng:2007jq} predicted Quark-Gluon-Plasma (QGP) or not,   is still a question of debate.
For long, strangeness enhancement is considered as a signature of QGP formation  \cite{Koch:1986ud}. In QGP environment, $gg\rightarrow s\bar{s}$ is abundant.
If not annihilated before hadronisation, early produced strange and anti-strange quarks will coalesce in to strange hadrons and 
compared to elementary pp collisions, strange particle production will be enhanced.  Recently, STAR   collaboration published their measurements of $\phi(s\bar{s})$ mesons in Au+Au \cite{Abelev:2007rw,:2008fd} and in Cu+Cu \cite{Collaboration:2008zk} collisions.  
Both in Au+Au and Cu+Cu collisions, compared to pp collisions, $\phi$ meson production is enhanced. 
However, it is uncertain whether or not the enhancement is due to increased production in QGP or due to canonical suppression of strangeness in pp collisions. STAR measurements of $\phi$ mesons in Au+Au collisions found to be compatible with a model based on recombination of thermal s quarks \cite{Hwa:2006vb}, strengthening the belief that in Au+Au collisions, a robust thermal parton source is created.

Relativistic hydrodynamics provides a convenient tool to analyse Au+Au collision data. It is assumed that in the collision a fireball is produced. Constituents of the fireball collide frequently to establish local thermal equilibrium sufficiently fast and after a certain time $\tau_i$, hydrodynamics become applicable. If the macroscopic properties of the fluid e.g. energy density, pressure, velocity etc. are known at the equilibration time $\tau_i$, the relativistic hydrodynamic equations can be solved to give the space-time evolution of the fireball till a given freeze-out condition such that interactions between the constituents are too weak to continue the evolution. 
Using suitable algorithm (e.g. Cooper-Frye) information at the freeze-out can be converted into particle spectra and can be directly compared with experimental data. Thus, hydrodynamics, in an indirect way, can characterize the initial condition of the medium produced in heavy ion collisions. Hydrodynamics equations are closed only with an equation of state (EOS) and one can investigate the possibility of phase transition in the medium. 
A host of experimental data produced in Au+Au collisions at RHIC, at c.m. energy $\sqrt{s}$=200 GeV, have been successfully analysed using ideal hydrodynamics \cite{QGP3}, with an equation of state with 1st order confinement-deconfinement phase transition. Multiplicity, mean $p_T$, $p_T$-spectra, elliptic flow etc.  of identified particles, are well explained in the ideal hydrodynamic model with QGP as the initial state.  Ideal hydrodynamics analysis of the RHIC data indicate that in central Au+Au collisions, at the equilibration time $\tau_i \approx$ 0.6 fm, 
  central energy density of the QGP fluid is $\varepsilon_i \approx$30 $GeV/fm^{-3}$ \cite{QGP3}. It may be mentioned that ideal hydrodynamics  description of data are not unblemished. 
  $p_T$ spectra or the elliptic flow are explained only up to  transverse momenta $p_T \approx 1.5 GeV$. At higher $p_T$ description deteriorates. Also ideal hydrodynamic description to data gets poorer in peripheral collisions.  
 
However, estimate of initial condition of the fluid can not be creditable unless dissipative effects are accounted for. 
Unlike in ideal fluid evolution, where initial and final state entropy remains the same, entropy is generated in viscous evolution. Consequently, to produce a fixed final state entropy,
viscous fluid require less initial energy density than an
ideal fluid. QGP viscosity is quite uncertain. Theoretical estimate cover a wide range, $\eta/s\approx$ 0-1. String theory based models (ADS/CFT) give a lower bound on viscosity of any matter $\eta/s \geq 1/4\pi$ \cite{Policastro:2001yc}. In a perturbative QCD, Arnold et al  \cite{Arnold:2000dr} estimated $\eta/s\sim$ 1. 
In a SU(3) gauge theory, Meyer \cite{Meyer:2007ic} gave the upper bound $\eta/s <$1.0, and his best estimate is $\eta/s$=0.134(33) at $T=1.165T_c$. 
At RHIC region, Nakamura and Sakai \cite{Nakamura:2005yf}
estimated the viscosity of a hot gluon gas  as $\eta/s$=0.1-0.4. Attempts have been made to estimate QGP viscosity directly from experimental data. 
Gavin and Abdel-Aziz \cite{Gavin:2006xd} proposed to measure viscosity from transverse momentum fluctuations. From the existing data on Au+Au collisions, they estimated that QGP viscosity as $\eta/s$=0.08-0.30. Experimental data on elliptic flow has also been used to estimate QGP viscosity. Elliptic flow scales with eccentricity. Departure form the scaling can be understood as due to off-equilibrium effect and utilised to estimate viscosity \cite{Drescher:2007cd} as, $\eta/s$=0.11-0.19. Experimental observation that elliptic flow scales with transverse kinetic energy is also used to estimate QGP viscosity, $\eta/s \sim$ 0.09 $\pm$ 0.015 \cite{Lacey:2006bc}, a value close to the ADS/CFT bound. From heavy quark energy loss, PHENIX collaboration \cite{Adare:2006nq} estimated
QGP viscosity $\eta/s\approx$ 0.1-0.16.

In recent years, considerable progress has been made in numerical implementation of dissipative hydrodynamics \cite{Teaney:2003kp,MR04,Koide:2007kw,Chaudhuri:2005ea,Heinz:2005bw,asis,Chaudhuri:2008sj,Romatschke:2007mq,Romatschke:2007jx,Baier:2006gy,Song:2007fn,Song:2007ux,Song:2008si}. 
 From the viscous hydrodynamic simulation of elliptic flow in  Au+Au collisions, 
Luzum and Romatschke \cite{Luzum:2008cw} obtained an upper bound to the ratio  $\eta/s <$ 0.4-0.5.  In \cite{Song:2008hj} Song and Heinz also argued that $\eta/s <$ 0.4 is a robust upper bound of QGP viscosity.
In a recent paper \cite{Chaudhuri:2009vx}, 
we have estimated QGP viscosity as $\eta/s\approx$0.25.   It was shown that $\phi$ mean $p_T$ is sensitive to QGP viscosity. STAR data \cite{Abelev:2007rw,:2008fd} on 
centrality dependence of $\phi$ mean $p_T$ in Au+Au collisions definitely reject ideal fluid or fluid with viscosity $\eta/s \leq$ 0.08-0.16. Data are explained only with $\eta/s$=0.25.  Initial central energy density of the fluid is $\varepsilon_i$=5.1 GeV, much less than that the estimated value $\sim$ 30 $GeV/fm^3$ in ideal hydrodynamics. Evolution of viscous fluid ($\eta/s$=0.25) also explains the centrality dependence of $\phi$ multiplicity, integrated $v_2$, $p_T$ spectra up to $\approx$ 3 GeV.


Purpose of the present paper is to show that viscosity over entropy ratio $\eta/s$=0.25, is consistent with the STAR measurements of $\phi$
elliptic flow in central and mid-central Au+Au collisions. It is also consistent with the recent STAR data on $\phi$ production in Cu+Cu collision at 200 GeV. The central energy density however, is less in Cu+Cu collisions, $\varepsilon_i$=3.48 $GeV/fm^3$.
The paper is organised as follows: in section \ref{sec2}, we briefly describe the 
hydrodynamical equations used to compute the evolution of ideal and viscous fluid. 
We have used a lattice motivated equation of state. Construction of the EOS
is also discussed in section \ref{sec2}. Simulation results   are discussed in section \ref{sec3}. Summary and conclusions are given in section \ref{sec4}.

\section{Hydrodynamical equations, equation of state and initial conditions}
\label{sec2}
\subsection{Hydrodynamical equations}
In the Israel-Stewart's theory of 2nd order dissipative hydrodynamics, space-time evolution of the fluid is obtained by solving,  
 
\begin{eqnarray}  
\partial_\mu T^{\mu\nu} & = & 0,  \label{eq3} \\
D\pi^{\mu\nu} & = & -\frac{1}{\tau_\pi} (\pi^{\mu\nu}-2\eta \nabla^{<\mu} u^{\nu>}) \nonumber \\
&-&[u^\mu\pi^{\nu\lambda}+u^\nu\pi^{\nu\lambda}]Du_\lambda. \label{eq4}
\end{eqnarray}

Eq.\ref{eq3} is the conservation equation for the energy-momentum tensor, $T^{\mu\nu}=(\varepsilon+p)u^\mu u^\nu - pg^{\mu\nu}+\pi^{\mu\nu}$, 
$\varepsilon$, $p$ and $u$ being the energy density, pressure and fluid velocity respectively. $\pi^{\mu\nu}$ is the shear stress tensor (we have neglected bulk viscosity and heat conduction). Eq.\ref{eq4} is the relaxation equation for the shear stress tensor $\pi^{\mu\nu}$.   
In Eq.\ref{eq4}, $D=u^\mu \partial_\mu$ is the convective time derivative, $\nabla^{<\mu} u^{\nu>}= \frac{1}{2}(\nabla^\mu u^\nu + \nabla^\nu u^\mu)-\frac{1}{3}  
(\partial . u) (g^{\mu\nu}-u^\mu u^\nu)$ is a symmetric traceless tensor. $\eta$ is the shear viscosity and $\tau_\pi$ is the relaxation time.  It may be mentioned that in a conformally symmetric fluid relaxation equation can contain additional terms  \cite{Song:2008si}.

\begin{figure}[t]
 \vspace{0.3cm} 
 \center
 \resizebox{0.35\textwidth}{!}{%
  \includegraphics{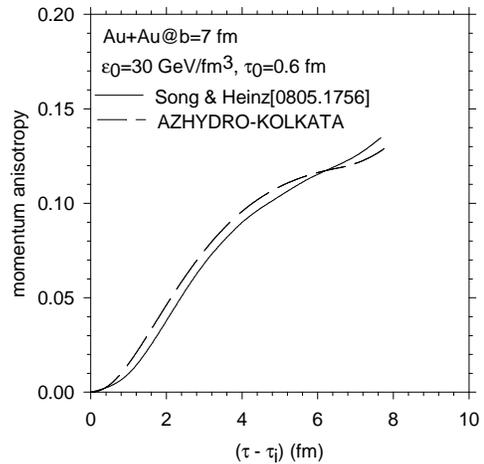}
}
\caption{Viscous fluid ($\eta/s$=0.08) simulation for temporal evolution of momentum anisotropy in b=7 fm Au+Au collision at RHIC. The solid line is the simulation result from VISH2+1 \cite{Song:2008si} 
and the dashed line is the simulation result from AZHYDRO-KOLKATA. Initial  condition of the fluid is very similar in both the simulations.}\label{F1}
\end{figure}

Assuming boost-invariance, Eqs.\ref{eq3} and \ref{eq4}  are solved in $(\tau=\sqrt{t^2-z^2},x,y,\eta_s=\frac{1}{2}\ln\frac{t+z}{t-z})$ coordinates, with a code 
  "`AZHYDRO-KOLKATA"', developed at the Cyclotron Centre, Kolkata.
 Details of the code can be found in \cite{Chaudhuri:2008sj}. To show that AZHYDRO-KOLKATA computes the evolution correctly, in Fig.\ref{F1}, we have compared the temporal evolution 
 of   momentum anisotropy $\varepsilon_p=\frac{<T^{xx}-T^{yy}>}{<T^{xx}+T^{yy}>}$ of a QGP fluid with a calculation
 of Song and Heinz \cite{Song:2008si}. Initial conditions are same for both the simulations.
Within 10\% or less, AZHYDRO-KOLKATA simulation  reproduces  Song and Heinz's  \cite{Song:2008si} result for temporal evolution of momentum anisotropy $\varepsilon_p$.

\subsection{Equation of state}
 
\begin{figure}[t]
\vspace{0.3cm} 
\center
 \resizebox{0.35\textwidth}{!}{%
  \includegraphics{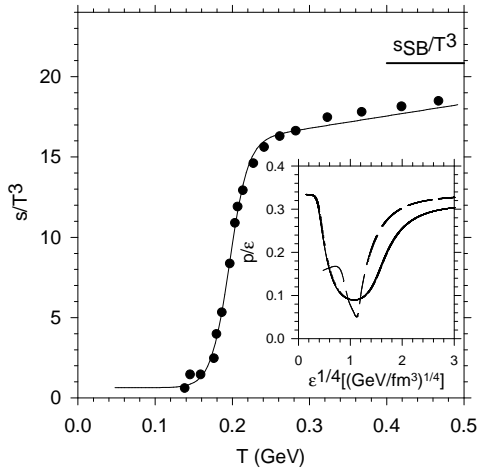}
}
\caption{Black circles are lattice simulation \cite{Cheng:2007jq} for entropy density. The black line is the parametric representation to the lattice simulations.
In the inset, the solid and dashed lines are the squared speed of sound in lattice based EOS and in an EOS incorporating 1st order transition \cite{QGP3}.}\label{F2}
\end{figure}

 One of the most important inputs of a hydrodynamic model  is the equation of state (EOS). Through this input macroscopic hydrodynamic models make contact with the microscopic world. 
Most of the hydrodynamical calculations are performed with EOS with a 1st order phase transition.
Huovinen    \cite{Huovinen:2005gy} reported an 'ideal' hydrodynamic simulation with 2nd order phase transition. He concluded that the experimental data (e.g. elliptic flow of proton or antiproton) are better explained with EOS with 1st order phase transition than with EOS with 2nd order phase transition.
However, lattice simulations \cite{Cheng:2007jq} indicate that confinement to deconfinement transition is a cross over, rather than a 1st or 2nd order phase transition. It is then essential that hydrodynamic simulations are done with EOS with cross-over transition rather than with EOS with 1st or 2nd order transition.  In Fig.\ref{F2},  a recent lattice simulation  \cite{Cheng:2007jq} for the  entropy density is
  shown.   The solid line in Fig.\ref{F2} is a parameterisation of the entropy density. 
  
\begin{equation}\label{eq1}
\frac{s}{T^3}=\alpha+[\beta+\gamma T][1+tanh\frac{T-T_c}{\Delta T}],
\end{equation}

From the parametric form of the entropy density, pressure and energy density can be obtained using the thermodynamic relations,

\begin{eqnarray}  
  p(T)&=&\int_0^T s(T) ds \label{eq2a} \\
  \varepsilon(T)&=&Ts -p \label{eq2b}.
  \end{eqnarray}

  Generally, in hydrodynamic simulations, hadronic phase is approximated by a (non-interacting) resonance hadron gas comprising all the resonances below 2-3 GeV. As the lattice simulation cover a wide temperature range below the cross over temperature, $T_{co}=196(3)$ MeV,
we choose to use the   lattice based EOS (Eq.\ref{eq1}-\ref{eq2b}) both in the QGP and in the hadronic phase. The idea is to expose the lattice simulation of EOS to experimental scrutiny. Indeed, lattice simulations are vague about the nature of the confined ($T < T_{co}$) phase. 
The confined phase is certainly unlike hadronic resonance gas. The  trace anomaly $(\varepsilon-3p)/T^4$ in the temperature range 140-200 MeV is approximately 30\% less than that of a hadronic resonance gas \cite{Cheng:2007jq}. One also note
 that at low temperature, effective degrees of freedom in the confined phase is $g_{h}\approx$2. In contrast, in hadronic resonance gas, $g_{h}\geq 40$.
In the inset of Fig.\ref{F2},  the squared speed of sound ($c_s^2\approx p/\varepsilon$) in the lattice based EOS  is compared with $c_s^2$ in an   EOS
with 1st order phase transition \cite{QGP3}, which model the quark phase with bag model, and the hadronic phase by the hadronic resonance gas. In 1st order EOS, $c_s^2$ fall sharply near the critical temperature. The fall is smoothened out in cross over  transition. Lattice based EOS is also softer. 
 
  \begin{figure}[t]
\vspace{0.3cm} 
\center
 \resizebox{0.35\textwidth}{!}{%
  \includegraphics{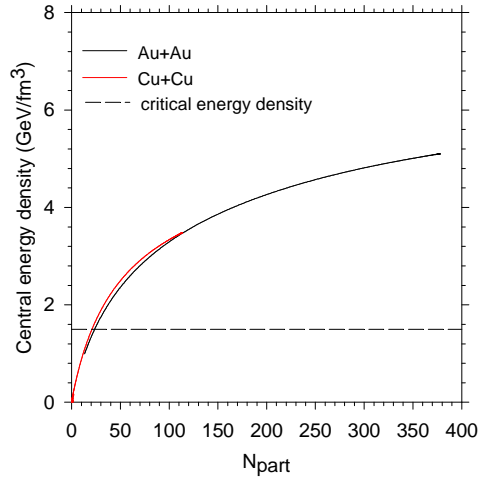}
}
\caption{(color online) The black and red solid lines are initial central energy density as a function of number of participants in Au+Au and Cu+Cu collisions.
The dashed line shows the critical energy density for the 
confinement-deconfinement cross-over transition.
 } \label{F3}
\end{figure} 
 
\subsection{Initial conditions}

Solution of partial differential equations (Eqs.\ref{eq3},\ref{eq4}) requires initial conditions, e.g.  transverse profile of the energy density ($\varepsilon(x,y)$), fluid velocity ($v_x(x,y),v_y(x,y)$) and shear stress tensor ($\pi^{\mu\nu}(x,y)$) at the initial time $\tau_i$. One also need to specify the viscosity ($\eta$) and the relaxation time ($\tau_\pi$). A freeze-out prescription is also needed to convert the information about fluid energy density and velocity to particle spectra and compare with experiment.

In \cite{Chaudhuri:2009vx}, we assumed that the fluid is thermalised at $\tau_i$=0.2 fm and the initial fluid velocity is zero, $v_x(x,y)=v_y(x,y)=0$.   Initial energy density was assumed to be distributed as \cite{QGP3}

\begin{equation} \label{eq6}
\varepsilon({\bf b},x,y)=\varepsilon_0[0.75 N_{part}({\bf b},x,y) +0.25 N_{coll}({\bf b},x,y)],
\end{equation}

\noindent
where b is the impact parameter of the collision. $N_{part}$ and $N_{coll}$ are the average participant and collision number respectively.
The
shear stress tensor was initialised with boost-invariant value. For the relaxation time, we used the Boltzmann estimate $\tau_\pi=3\eta/4p$.
The freeze-out was fixed at $T_F$=150 MeV. In Eq.\ref{eq6},  
$\varepsilon_0$ is a parameter which does not depend on the impact parameter of the collision. Assuming that $\eta/s$ remain a constant throughout the evolution, for a set of values $\eta/s$=0 (ideal fluid), 0.08, 0.16 and 0.25, we fit $\varepsilon_0$  to reproduce STAR measurements of $\phi$ multiplicity in 0-5\% centrality Au+Au collisions. Centrality dependence of $\phi$ multiplicity and $\phi$ mean $p_T$ are  simultaneously explained only with viscosity over entropy ratio $\eta/s$=0.25. The corresponding central energy density, in b=0 Au+Au collision is $\varepsilon_i$=5.1 $GeV/fm^3$.

  \begin{figure}[t]
\vspace{0.3cm} 
\center
 \resizebox{0.35\textwidth}{!}{%
  \includegraphics{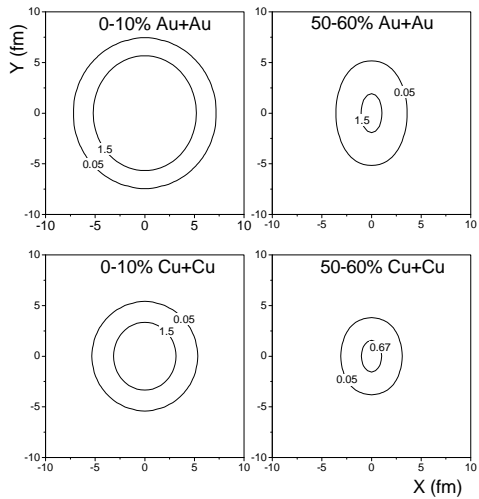}
}
\caption{The top two panels show the energy density contours in 0-10\% and 50-60\% centrality Au+Au collisions. The contours are drawn at $\varepsilon$=1.5 and 0.05 $GeV/fm^3$. They correspond to confinement-deconfinement cross-over and freeze-out respectively. The bottom two panels show the energy density contours in Cu+Cu collisions. In 50-60\% centrality Cu+Cu collisions, initially QGP is not produced.
 } \label{F4}
\end{figure}

  For Cu+Cu collisions,
we only change the central energy density, other parameters remain unchanged.
The initial energy density in Cu+Cu collisions is obtained by fitting $\phi$ multiplicity in central 0-10\% Cu+Cu collisions. The fitted value corresponds to central energy density $\varepsilon_i$=3.48 $GeV/fm^3$ in  b=0 Cu+Cu collisions. As it will be shown below, QGP fluid, initialised with $\varepsilon_i$=3.48 $GeV/fm^3$ reproduces most of the STAR measurements on $\phi$ mesons in Cu+Cu collisions. 
It is interesting to note that $\varepsilon^{au}_i/\varepsilon^{cu}_i \approx A^{1/3}_{au}/A^{1/3}_{cu}$. Apparently, initial energy density scales with nuclear radius. With boost-invariance, the initial system is a cylinder of radius $R=r_0A^{1/3}$, infinitely extended in the rapidity direction. 
Phenomenological relation $\varepsilon^{au}_i/\varepsilon^{cu}_i \approx A^{1/3}_{au}/A^{1/3}_{cu}$ indicate that 
at the freeze-out also the system can be approximated again by a cylinder with radius $R^\prime \propto R$.

Before we compare hydrodynamic simulation for $\phi$ production with experiment, it is interesting to compare initial central energy density in different centrality rages of Au+Au and Cu+Cu collisions. 
In Fig.\ref{F3}, we have compared the initial central energy density   in Au+Au and in Cu+Cu collisions as a function of participant number. The black and red lines are for Au+Au and  Cu+Cu  collisions respectively. In the region where they overlap, initial central energy density in Au+Au and Cu+Cu collisions are similar. One then expects that $\phi$ spectra in mid central Au+Au and central Cu+Cu collisions will be similar.
The expectation is fulfilled in STAR experiment (e.g. $\phi$ $p_T$ spectra in 40-50\% Au+Au and 10-20\% Cu+Cu collisions are nearly identical).
In Fig.\ref{F3}, the  dashed line is the energy density ($\varepsilon_{co} \approx$1.5 $GeV/fm^3$) for the confinement-deconfinement 
 cross-over.   In most of the collisions, in the central region, initially the fluid is produced in the deconfined state. However, energy density has a distribution, fluid in the central region is at higher density than the fluid at periphery. Thus in mid-central collisions, only a small portion of the fluid will be in the deconfined phase. In Fig.\ref{F4}, in four panels, we have shown the contours of initial  energy density  in 0-10\% and 50-60\% centrality Au+Au and in Cu+Cu collisions. Contours are drawn at $\varepsilon_{co}$=1.5 $GeV/fm^3$ and $\varepsilon_{fo}$=0.05 $GeV/fm^3$, corresponding to cross-over energy density and freeze-out.  In 0-10\% centrality Au+Au and Cu+Cu collisions, initially, fluid in the central
region is in QGP state. But fraction of fluid in QGP state is larger in Au+Au than in Cu+Cu collisions. One can immediately say that hard probe signature of QGP formation will be less prominent in 0-10\% centrality Cu+Cu collisions than in 0-10\% centrality Au+Au collisions. For example, one  can 
conjecture that in 0-10\% Cu+Cu collisions $J/\psi$'s will be less suppressed than in a 0-10\% centrality Au+Au collision. Experiments do vindicate the conjecture \cite{Adler:2003rc,Adare:2006ns,Adare:2008sh} .  

 \begin{figure}[t]
 \vspace{0.3cm} 
 \center
 \resizebox{0.35\textwidth}{!}{%
  \includegraphics{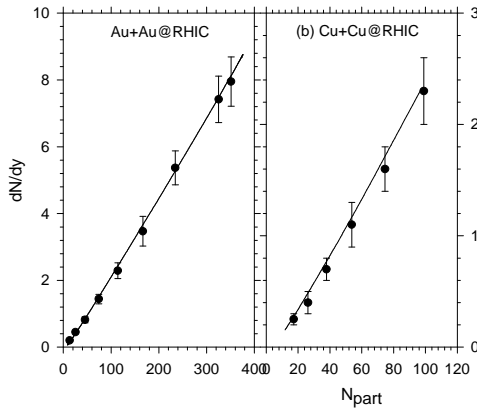}
}
\caption{ (a) Filled circles are STAR data on the centrality dependence of $\phi$ meson multiplicity.
The black line is viscous ($\eta/s$=0.25) hydrodynamic fit to the data. The initial time $\tau_i$=0.2 fm, initial central energy density $\varepsilon_i$=5.1 $GeV/fm^3$, freeze-out temperature is $T_F$=150 MeV. (b) same as in (a) but for Cu+Cu collisions. The initial energy density is $\varepsilon_i$=3.48 $GeV/fm^3$.  }\label{F5}
\end{figure}
 
\section{Results}\label{sec3}

\subsection{centrality dependence of $\phi$ multiplicity, mean $p_T$ and integrated $v_2$ in Au+Au and Cu+Cu collisions}

In \cite{Chaudhuri:2009vx}, we have shown the viscous ($\eta/s$=0.25) hydrodynamics fit to the STAR data on $\phi$ multiplicity, mean $p_T$ and integrated flow in Au+Au collisions. For completeness purpose, here also, we show the fits along with the fit obtained to Cu+Cu data.
In Figs.\ref{F5}a and \ref{F5}b, the STAR measurements for the centrality dependence of $\phi$ multiplicity ($dN/dy$) in Au+Au \cite{Abelev:2007rw,:2008fd} and Cu+Cu \cite{Collaboration:2008zk} collisions are shown.
In the region where $N_{part}$ overlap, $\phi$ multiplicity is nearly identical in Au+Au and in Cu+Cu collisions.  In Fig.\ref{F5}a and ,b
the solid lines are the hydrodynamic predictions for $\phi$ multiplicity.  Evolution of viscous ($\eta/s$=0.25) QGP fluid thermalised at $\tau_i$=0.2 fm and initialised with central energy density 5.1 $GeV/fm^3$ in Au+Au collisions and  3.48 $GeV/fm^3$ in Cu+Cu collisions, reproduces the data in all the centrality ranges of collisions. 
As indicated above, we have used only the most central collision data (0-5\% in case of Au+Au collisions and 0-10\% in case of Cu+Cu collisions) to fix the initial energy density. Glauber model initial condition (Eq.\ref{eq6}) correctly incorporate the centrality dependence and $\phi$ multiplicity is reproduced in all the centrality ranges of collisions.
 \begin{figure}[t]
\vspace{0.3cm} 
\center
 \resizebox{0.35\textwidth}{!}{%
  \includegraphics{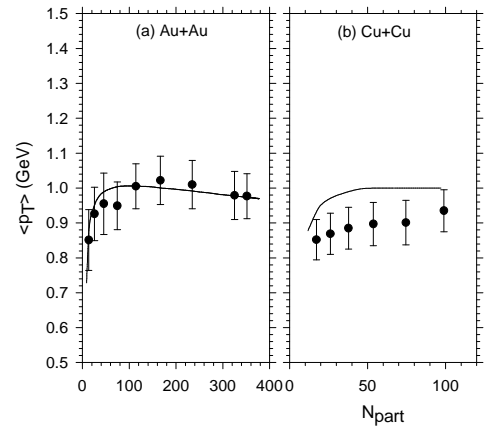}
}
\caption{(a) Filled circles are STAR data \cite{Abelev:2007rw} on centrality dependence of mean $p_T$ of $\phi$ mesons in Au+Au collisions. The black line is the fit 
obtained to the data in viscous hydrodynamics. (b) same as in (a) but for Cu+Cu collisions.}
\label{F6}
\end{figure} 

In Fig.\ref{F6}a and b, we have shown the STAR measurements of$\phi$  mean $p_T$ in Au+Au \cite{Abelev:2007rw,:2008fd} and Cu+Cu \cite{Collaboration:2008zk} collisions.  Within the error bars, centrality dependence of $\phi$ mean $p_T$ in Au+Au and in Cu+Cu collisions are nearly identical (though central value is consistently higher in Au+Au collisions).  
The black lines in Fig.\ref{F6}a,b are fit to the data in viscous hydrodynamics. In viscous hydrodynamics also, $\phi$ mean $p_T$ do not show any appreciable dependence on system size.  STAR measurements of mean $p_T$ in Au+Au collisions are nicely reproduced.   $\phi$ mean $p_T$ in Cu+Cu are reproduced within 10\% or less. 

\begin{figure}[t]
\vspace{0.3cm}
\center
 \resizebox{0.35\textwidth}{!}{%
  \includegraphics{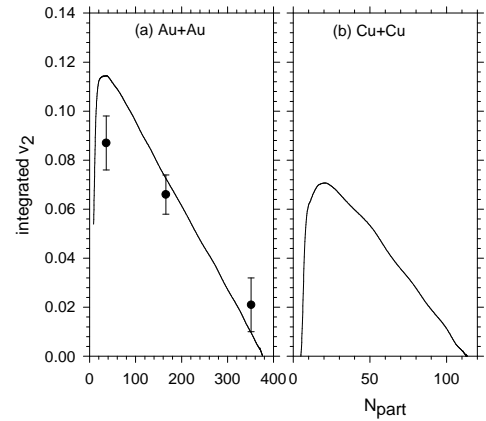}
}
\caption{ (a) filled circles are STAR data on integrated elliptic flow in Au+Au collisions. The black line shows the centrality dependence of integrated $v_2$.
(b) centrality dependence of integrated $v_2$ in Cu+Cu collisions.}\label{F7}
\end{figure}

Black lines in Fig.\ref{F7}a and b, are the viscous hydrodynamics predictions for the centrality dependence of integrated $v_2$ in Au+Au and Cu+Cu collisions. 
In Au+Au collisions, STAR measured integrated $v_2$ in 0-5\%, 10-40\% and 40-80\% centrality collisions \cite{Abelev:2007rw,:2008fd}. Filled circles in Fig.\ref{F7}a are the STAR measurements. Except for the very 
peripheral collision, hydrodynamic prediction is agreement with STAR data.
In Cu+Cu collisions, integrated $v_2$ is not measured yet. However, simulation results indicate that compared to Au+Au collisions, $\phi$ meson integrated flow is less in Cu+Cu collisions. The reason is understood. Initial eccentricity is small in Cu+Cu than in Au+Au collisions.
Elliptic flow has size dependence, smaller the system, less is the flow.  

\subsection{$\phi$ $p_T$-spectra in Au+Au and Cu+Cu collisions}

\begin{figure}[t]
\vspace{0.3cm} 
\center
 \resizebox{0.35\textwidth}{!}{%
  \includegraphics{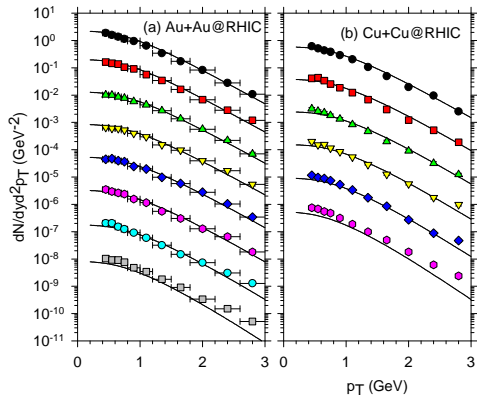}
}
\caption{(color online) (a) STAR data \cite{Abelev:2007rw} $\phi$ meson $p_T$ spectra in 0-5\%, 0-10\%, 10-20\%, 20-30\%, 30-40\%, 40-50\%, 50-60\% and 60-70\% centrality Au+Au collisions. The black lines are $\phi$ spectra from evolution of viscous fluid.  
(b) STAR data on $p_T$ spectra of $\phi$ in Cu+Cu collisions in 0-10\%, 10-20\%, 20-30\%, 30-40\%, 40-50\% and 50-60\% Cu+Cu collisions.
The black lines are viscous hydrodynamic predictions.}
 \label{F8}
\end{figure}

STAR measurements \cite{Abelev:2007rw} for $\phi$ meson $p_T$ spectra in
 0-5\%, 5-10\%, 10-20\%,20-30\%, 30-40\%, 40-50\%, 50-60\% and 60-70\%  centrality Au+Au collisions are shown in Fig.\ref{F8}a. As noted by the STAR collaboration  \cite{Abelev:2007rw} $\phi$ $p_T$-spectra
up to 30-40\% centrality collisions are well fitted by an exponential, indicating thermal production of $\phi$ in central collisions. Exponential fit worsen in more peripheral collisions. A Levy function
(which has an exponential shape at low $p_T$ and power law shape at large $p_T$) fits the peripheral data. Apparently, in peripheral collisions non-thermal source contribute to $\phi$ meson production.  
The black lines in Fig.\ref{F8}a are 
$p_T$ spectra from evolution of viscous QGP fluid. Except for very peripheral collisions, data up to $p_T$=3 GeV are well explained in viscous hydrodynamics. 
At larger $p_T$ (not shown in Fig.\ref{F8}), viscous hydrodynamics under predict the $p_T$-spectra. At large $p_T$, other sources e.g. pQCD processes can contribute and hydrodynamic models may not be reliable.  

STAR collaboration recently published their measurements for $\phi$ $p_T$-spectra in Cu+Cu collisions \cite{Collaboration:2008zk}.
STAR measurements \cite{Collaboration:2008zk} for $\phi$ $p_T$ spectra in 0-10\%, 10-20\%, 10-20\%, 20-30\%, 30-40\%, 40-50\% and 50-60\% centrality Cu+Cu collisions are shown in Fig.\ref{F8}b. A Levy function also fits the $p_T$ spectra in Cu+Cu collisions \cite{Collaboration:2008zk}. For similar $N_{part}$, $\phi$ spectra in Au+Au and Cu+Cu collisions are similar. The parameters of the Levy function in Au+Au collision and Cu+Cu collision are also similar for nearly identical for  participant numbers. 
Black lines in Fig.\ref{F8}b are the predictions from viscous hydrodynamics.
Here again, except for the very peripheral   (50-60\%) collisions, data, up to $p_T$=3 GeV  are well explained. The results indicate that $\phi$ $p_T$ spectra,
up to $p_T$= 3GeV,
both in Au+Au and in Cu+Cu collisions are consistent with hydrodynamic evolution
of QGP fluid with viscosity over entropy ratio $\eta/s$=0.25. The central energy density of the fluid in b=0 Au+Au (Cu+Cu) collisions is $\approx$5.1(3.48) $GeV/fm^3$  . 


\subsection{$N(\Omega)/N(\phi)$ vs. $p_T$}

\begin{figure}[t]
 \center
 \resizebox{0.35\textwidth}{!}{%
  \includegraphics{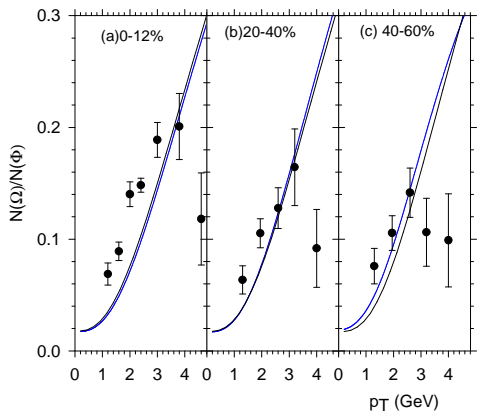}
}
\caption{(color online) Filled circles are STAR data on the $p_T$ dependence of the ratio $N(\Omega)/N(\phi)$ in Au+Au collisions in 0-12\%, 20-40\% and 40-80\%
centrality Au+Au collisions.  The black lines are the viscous hydrodynamic predictions for the ratio in Au+Au collisions. The blue lines are the predictions for the ratio in Cu+Cu collisions.}
   \label{F9}
\end{figure}

STAR collaboration measured the transverse momentum dependence of the ratio $N(\Omega)/N(\phi)$ in Au+Au collisions \cite{Abelev:2007rw}.  STAR measurements of the ratio in 
 0-12\%, 20-40\% and 40-60\% centrality Au+Au collisions are shown in  the   three panels, a,b,c of Fig.\ref{F9}.  
 The ratio increases with $p_T$ till $p_T\approx$ 3-4 GeV then drops. In peripheral collisions, the ratio drops at lower $p_T$ than in more central collisions.  
Both $\Omega(sss)$ and $\phi(s\bar{s})$ are strange particles, devoid of any non-strange quarks. The ratio $N(\Omega)/N(\phi)$ can shed light on the production mechanism of strange particles, specifically, strange baryon and
mesons. The ratio can also test a model.  Correct reproduction of the ratio will indicate that the strangeness sector is correctly modeled.  
 Models based on recombination of thermal strange quarks \cite{Hwa:2006vb} can reproduce the $\phi$ meson $p_T$ spectra up to $p_T$= 5 GeV. The model \cite{Hwa:2006vb} also reproduces the ratio $N(\Omega)/N(\phi)$ up to $p_T\approx$ 4 GeV,  reproduces the decreasing trend at $p_T>$ 4 GeV. But at $p_T >$ 4 GeV, the model largely over predict the ratio, indicating that at large $p_T$,
in recombination models $\Omega$'s are more produced than in experiment.
In Fig.\ref{F9}, the black lines are hydrodynamic predictions for the ratio in Au+Au collisions. Hydrodynamics predictions are shown up to $p_T$= 5 GeV.
  The ratio increases with $p_T$, and continue to increase even at large $p_T$.   
     Evolution of viscous QGP do not reproduce the experimental trend that the ratio decreases beyond a certain $p_T$. It is not expected also. As noted earlier, $\phi$ spectra at $p_T>$ 3 GeV are not reproduced.
In $p_T$ range $p_T \leq$ 3 GeV, 
viscous hydrodynamics appear to under predict the ratio in $p_T$ range $p_T \leq$ 3 GeV. For example in 0-12\% centrality collisions, the ratio is under predicted by $\sim$ 40\%. In more peripheral collisions, the ratio is less under predicted.
The result is interesting. $\phi$ $p_T$-spectra, up to $p_T$=3 GeV, are well reproduced in viscous hydrodynamics. Apparently, in viscous hydrodynamics, $\Omega$'s are not produced in sufficient number. 
The ratio $N(\Omega)/N(\phi)$ is not measured in Cu+Cu collisions. The blue lines in Fig.\ref{F9} are the predictions for the ratio in Cu+Cu collisions.
 The ratio is nearly identical to that in Au+Au collisions. The ratio $N(\Omega)/N(\phi)$ do not show any system size dependence.

\subsection{$\phi$   elliptic flow in Au+Au and Cu+Cu collisions}

\begin{figure}[t]
 \vspace{0.3cm} 
 \center
 \resizebox{0.35\textwidth}{!}{%
  \includegraphics{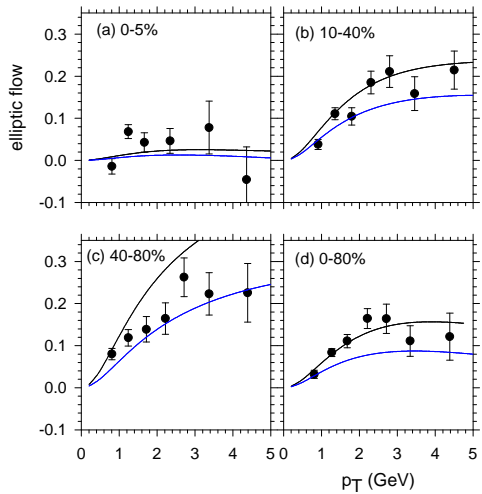}
}
\caption{(color online) Filled circles in panels a,b,c and d are STAR measurements for $\phi$ meson  elliptic flow in 0-10\%, 10-40\% and 40-80\% and 0-80\% centrality Au+Au collisions. The black lines in the figure are the hydrodynamic model predictions for  elliptic flow in Au+Au collisions. 
Blue lines are the predicted flow in Cu+Cu collisions.}\label{F10}
\end{figure}

STAR collaboration measured $\phi$ meson elliptic flow in 0-5\%, 10-40\% 
and 40-80\% and 0-80\% (minimum bias)  centrality Au+Au collisions \cite{Abelev:2007rw}. STAR measurements for the elliptic flow are shown in Fig.\ref{F10}.
In Fig.\ref{F10}, the black lines are the elliptic flow from evolution of viscous fluid with central energy density $\varepsilon_0$=5.1 $GeV/fm^3$. Even though as mentioned earlier, viscous hydrodynamics is not reliable beyond $p_T$=3 GeV ($\phi$ $p_T$-spectra are under predicted), we have shown predictions for flow up to   $p_T$=5 GeV.
 In 0-5\% collisions, elliptic flow is very small, and   viscous fluid evolution reproduces the flow. Elliptic flow in 10-40\% collisions is also reproduced in the model. However, flow is largely over predicted in 40-80\% centrality collisions.
40-80\% centrality collisions approximately corresponds to b=11 fm Au+Au collision. Hydrodynamic models are not reliable at 
 such peripheral collisions. Interestingly, elliptic flow in 0-80\% centrality collisions is also well reproduced in viscous hydrodynamics.   Considering that only the central energy density is fixed to reproduce $\phi$ multiplicity in 0-5\% centrality Au+Au collisions, reproduction of $\phi$ $p_T$ spectra and elliptic flow in central and mid central collisions can be considered as a great success of viscous hydrodynamics.  

Blue lines in Fig.\ref{F10}, are the predictions for flow in   0-5\%, 10-40\%  and 40-80\% and minimum bias Cu+Cu collisions. Compared to Au+Au collisions, elliptic flow is $\sim$ 10\% less in Cu+Cu collisions. It is consistent with our predictions for integrated $v_2$. Integrated $v_2$ is also less in Cu+Cu than in Au+Au collisions (see Fig.\ref{F7}).  
$\phi$ meson elliptic flow in Cu+Cu collisions are not measured yet and
the predictions   can not be tested against experiment. Future experiments can verify the predictions.
  
Present analysis indicate that QGP fluid, with viscosity over entropy ratio $\eta/s$=0.25, is consistent with most of the published STAR data on 
$\phi$ production in central and mid-central Au+Au and Cu+Cu collisions. The initial central energy density of the  fluid is $\approx$5.1 $GeV/fm^3$ in Au+Au collisions and $\approx$ 3.48 $GeV/fm^3$ in Cu+Cu collisions.  
%
%
Present estimate of viscosity and initial energy density are obtained by fitting experimental data in a hydrodynamic model. Limitations of the model must be discussed. We have neglected bulk viscosity.  In general bulk viscosity is much smaller than shear viscosity.  However, recent lattice simulation \cite{Cheng:2007jq} indicate that trace anomaly is non-zero near the cross-over temperature. Using the lattice data, Kharzeev et al \cite{Kharzeev:2007wb} computed bulk viscosity of QGP. Near the cross over temperature, bulk viscosity can be significantly large. 
Experimental data include effect of both the shear and bulk viscosity. Neglecting bulk viscosity will result into overestimating the shear viscosity.
Then $\eta/s$=0.25 is an upper bound of QGP viscosity.
However, since bulk viscosity is appreciable only near the cross over temperature, we do not expect substantial entropy production due to bulk viscosity and estimate of shear viscosity will largely remain unaltered.

\section{Summary and conclusions} \label{sec4} 
 
To summarise, in the Israel-Stewart's theory of dissipative hydrodynamics, we have simulated $\phi$ production 
from Au+Au and Cu+Cu collisions at $\sqrt{s}$=200 GeV.  
In an earlier publication \cite{Chaudhuri:2009vx}, we have shown that the STAR data on $\phi$ mean $p_T$
in Au+Au collisions is sensitive to viscosity and estimated QGP viscosity as $\eta/s$=0.25. For $\eta/s$=0.25, QGP fluid, thermalised at $\tau_i$=0.2 fm and initialised with central energy density 5.1 $GeV/fm^3$, explain centrality dependence of $\phi$ mean $p_T$, multiplicity, integrated $v_2$ and $p_T$ spectra (up to $p_T\approx$ 3 GeV). It is now shown that the STAR data on $\phi$ elliptic flow in central and mid-central Au+Au collisions are also explained in evolution of QGP fluid with viscosity to entropy ratio $\eta/s$=0.25.   $\eta/s$=0.25 is also consistent with STAR data on $\phi$ meson $p_T$ spectra in Cu+Cu collisions. In Cu+Cu collisions, central energy density is 3.48 $GeV/fm^3$, $\sim$1.5 times less that in Au+Au collisions. We have given predictions for elliptic flow in Cu+Cu collisions. Predicted elliptic flow in Cu+Cu collisions is $\sim$10\% less than that in Au+Au collisions. In conclusion, STAR data on intermediate range $p_T\leq$ 3 GeV, $\phi$ meson production is Au+Au and Cu+Cu collisions is consistent with hydrodynamic evolution of QGP fluid with viscosity to entropy ratio   $\eta/s\approx$0.25.


\begin{thebibliography}{99}
\bibitem{BRAHMSwhitepaper}
 BRAHMS Collaboration, I. Arsene {\it et al.},  
Nucl. Phys. A {\bf 757}, 1 (2005). 
 
\bibitem{PHOBOSwhitepaper} 
PHOBOS Collaboration,  B. B. Back {\it et al.},  
Nucl. Phys. A {\bf 757}, 28 (2005). 
 
\bibitem{PHENIXwhitepaper} 
PHENIX Collaboration, K.~Adcox {\it et al.}, 
Nucl. Phys. A {\bf 757} 184 (2005).  
  
\bibitem{STARwhitepaper} 
STAR Collaboration, J. Adams {\it et al.}, 
Nucl. Phys. A {\bf 757} 102 (2005).

\bibitem{lattice} 
Karsch F, Laermann E, Petreczky P, Stickan S and Wetzorke I, 
2001 {\it Proccedings of NIC Symposium} (Ed. H. Rollnik and D. Wolf, John 
von Neumann Institute for Computing, J\"ulich, NIC Series, vol.9, 
ISBN 3-00-009055-X, pp.173-82,2002.)

\bibitem{Cheng:2007jq}
  M.~Cheng {\it et al.},
  Phys.\ Rev.\  D {\bf 77}, 014511 (2008)
  [arXiv:0710.0354 [hep-lat]].
  
\bibitem{Koch:1986ud}
  P.~Koch, B.~Muller and J.~Rafelski,
  Phys.\ Rept.\  {\bf 142}, 167 (1986).
  
\bibitem{Abelev:2007rw}
  B.~I.~Abelev {\it et al.}  [STAR Collaboration],
  Phys.\ Rev.\ Lett.\  {\bf 99}, 112301 (2007)
  [arXiv:nucl-ex/0703033].
  
\bibitem{:2008fd}
   {\it et al.}  [STAR Collaboration],
  arXiv:0809.4737 [nucl-ex].

\bibitem{Collaboration:2008zk}
  S.~Collaboration,
  arXiv:0810.4979 [nucl-ex].
  
\bibitem{Hwa:2006vb}
  R.~C.~Hwa and C.~B.~Yang,
  arXiv:nucl-th/0602024.
  
  
  
  
  
  
  
  
  
  
\bibitem{QGP3}
P.~F. Kolb and U. Heinz,
in {\it Quark-Gluon Plasma 3}, edited by R.~C. Hwa and 
X.-N. Wang (World Scientific, Singapore, 2004), p.~634.

\bibitem{Policastro:2001yc}
  G.~Policastro, D.~T.~Son and A.~O.~Starinets,
  Phys.\ Rev.\ Lett.\  {\bf 87}, 081601 (2001).
  
\bibitem{Arnold:2000dr}
  P.~Arnold, G.~D.~Moore and L.~G.~Yaffe,
  JHEP {\bf 0011}, 001 (2000),JHEP {\bf 0305}, 051 (2003).
\bibitem{Meyer:2007ic}
  H.~B.~Meyer,
  Phys.\ Rev.\  D {\bf 76}, 101701 (2007)
  [arXiv:0704.1801 [hep-lat]].
\bibitem{Nakamura:2005yf}
  A.~Nakamura and S.~Sakai,
  Nucl.\ Phys.\  A {\bf 774}, 775 (2006).
  
\bibitem{Gavin:2006xd}
  S.~Gavin and M.~Abdel-Aziz,
  Phys.\ Rev.\ Lett.\  {\bf 97}, 162302 (2006)
  [arXiv:nucl-th/0606061].
  
\bibitem{Drescher:2007cd}
  H.~J.~Drescher, A.~Dumitru, C.~Gombeaud and J.~Y.~Ollitrault,
  Phys.\ Rev.\  C {\bf 76}, 024905 (2007)
  [arXiv:0704.3553 [nucl-th]].
  
\bibitem{Lacey:2006bc}
  R.~A.~Lacey {\it et al.},
  Phys.\ Rev.\ Lett.\  {\bf 98}, 092301 (2007)
  [arXiv:nucl-ex/0609025].
  
\bibitem{Adare:2006nq}
  A.~Adare {\it et al.}  [PHENIX Collaboration],
  Phys.\ Rev.\ Lett.\  {\bf 98}, 172301 (2007)
  [arXiv:nucl-ex/0611018].
  
\bibitem{Teaney:2003kp}
  D.~Teaney,
  Phys.\ Rev.\  C {\bf 68}, 034913 (2003)
  [arXiv:nucl-th/0301099].

\bibitem{MR04}  A.~Muronga and D.~H.~Rischke,
  nucl-th/0407114\,(v2).
\bibitem{Koide:2007kw}
  T.~Koide, G.~S.~Denicol, Ph.~Mota and T.~Kodama,
  Phys.\ Rev.\  C {\bf 75}, 034909 (2007).

\bibitem{Chaudhuri:2005ea}
  A.~K.~Chaudhuri and U.~W.~Heinz,
  J.\ Phys.\ Conf.\ Ser.\  {\bf 50}, 251 (2006).

\bibitem{Heinz:2005bw}
  U.~W.~Heinz, H.~Song and A.~K.~Chaudhuri,
  Phys.\ Rev.\  C {\bf 73}, 034904 (2006).

\bibitem{asis}
  A.~K.~Chaudhuri,
  Phys.\ Rev.\  C {\bf 74}, 044904 (2006).
  arXiv:nucl-th/0703027;
  arXiv:nucl-th/0703029;
  arXiv:0704.0134 [nucl-th].
 
\bibitem{Chaudhuri:2008sj} A.~K.~Chaudhuri,
 arXiv:0801.3180 [nucl-th].

\bibitem{Romatschke:2007mq}
  P.~Romatschke and U.~Romatschke,
  Phys.\ Rev.\ Lett.\  {\bf 99}, 172301 (2007)
  [arXiv:0706.1522 [nucl-th]].
\bibitem{Romatschke:2007jx}
  P.~Romatschke,
  Eur.\ Phys.\ J.\  C {\bf 52}, 203 (2007)
  [arXiv:nucl-th/0701032].
\bibitem{Baier:2006gy}
  R.~Baier and P.~Romatschke,
  Eur.\ Phys.\ J.\  C {\bf 51}, 677 (2007)
  [arXiv:nucl-th/0610108].
\bibitem{Song:2007fn}
  H.~Song and U.~W.~Heinz,
  arXiv:0709.0742 [nucl-th].
\bibitem{Song:2007ux}
  H.~Song and U.~W.~Heinz,
  arXiv:0712.3715 [nucl-th].
 
\bibitem{Song:2008si}
  H.~Song and U.~W.~Heinz,
  Phys.\ Rev.\  C {\bf 78}, 024902 (2008)
  [arXiv:0805.1756 [nucl-th]].
  
\bibitem{Luzum:2008cw}
  M.~Luzum and P.~Romatschke,
  Phys.\ Rev.\  C {\bf 78}, 034915 (2008)
  [arXiv:0804.4015 [nucl-th]].
\bibitem{Song:2008hj}
  H.~Song and U.~W.~Heinz,
  arXiv:0812.4274 [nucl-th].
  
  
\bibitem{Chaudhuri:2009vx}
  A.~K.~Chaudhuri,
  arXiv:0901.0460 [nucl-th].


   
\bibitem{Huovinen:2005gy}
  P.~Huovinen,
  Nucl.\ Phys.\  A {\bf 761}, 296 (2005)
  [arXiv:nucl-th/0505036].

  
  
\bibitem{Adler:2003rc}
  S.~S.~Adler {\it et al.}  [PHENIX Collaboration],
  Phys.\ Rev.\  C {\bf 69}, 014901 (2004)
  [arXiv:nucl-ex/0305030].
\bibitem{Adare:2006ns}
  A.~Adare  [PHENIX Collaboration],
Phys.\ Rev.\ Lett.\  {\bf 98}, 232301 (2007)
  [arXiv:nucl-ex/0611020].
\bibitem{Adare:2008sh}
  A.~Adare {\it et al.}  [PHENIX Collaboration],
  arXiv:0801.0220 [nucl-ex].
  
  
\bibitem{Kharzeev:2007wb}
  D.~Kharzeev and K.~Tuchin,
  JHEP {\bf 0809}, 093 (2008)
  [arXiv:0705.4280 [hep-ph]].


\end{thebibliography}
\end{document}